\documentstyle[preprint,pra,aps]{revtex}
\begin{document}
\title{\bf
Renormalization of the one-body off-diagonal Coulomb field in nuclei
}
\author{N.Auerbach and O.K.Vorov}
\address{
School of Physics and Astronomy,
Tel Aviv University,
Ramat Aviv, 69978 Tel Aviv, Israel
}
\date{June 17, 1997}
\maketitle
\begin{abstract}
Isospin violation effects in nuclei are 
investigated within a microscopic approach 
which takes into account the influence of the residual strong interaction
on the Coulomb interaction.
The renormalization of the off-diagonal Coulomb 
field acting within a nucleus,
by the residual strong interaction is
calculated analytically in a simplified RPA.
From the expression for the 
suppression coefficient of the isospin violating
part of the Coulomb field, 
the isospin violating
spreading widths of  
isobaric analog states are derived.
The resulting reduction of the width is in agreement
with the 
data on the isospin symmetry restoration 
and with some earlier
evaluations of the isospin violating spreading widths.
\end{abstract}
\pacs{PACS: }
The off-diagonal part of the Coulomb potential in nuclei plays an important
role in determining some properties of nuclear structure 
\cite{AK}-\cite{HAW}.
Many effects of isospin mixing including the widths of isobaric analog 
resonances are governed by off-diagonal matrix elements of the Coulomb 
one-body potential. 
It was realized \cite{AK}-\cite{A-PLB}
that the off-diagonal one-body Coulomb matrix 
elements are strongly renormalized by the strong interaction in nuclei.
This renormalization leads to reduced isospin mixing and the isospin symmetry 
is preserved to a better degree than what one would expect 
from unrenormalized Coulomb interactions. The method in which 
the renormalization was treated is by introducing 
collective effects into the radial 
motion in nuclei.
The notion of the collective giant isovector monopole
state (IVM) was introduced and its role as a mediator 
of many Coulomb mixing effects 
\cite{AK}-\cite{BM}
was discussed 
twenty five years ago.
The reduction in the Coulomb matrix elements is achieved by introducing 
a collective RPA upwards shift of the IVM 
and by a reduction of its strength due to the repulsive nature 
of the particle-hole interaction in the isovector channel.

The problem of isospin symmetry violation in complex nuclear
states attracted attention for many years \cite{AK,A,GW,HAW,MBES,AG}
in the
general context of theory of chaotic systems \cite{BRODY}.
The case of isospin symmetry plays a special role in this respect because
the very microscopic origin of its violation is well known
and therefore constitutes a good example
where one can test the theories
of chaos in many-body systems.

A natural way to achieve this is to use an explicit realistic 
strong interaction that describes well nuclear properties 
and to employ RPA to evaluate the one-body Coulomb matrix elements.
We are able to develop approximate analytical formulas
which describe the renormalization of the off-diagonal Coulomb potential 
due to the strong nuclear force acting in nuclei.
When realistic nuclear forces are used,
the resulting Coulomb matrix elements are quenched 
and and are of the size that is capable to 
explain the experimentally observed reduction of the
isospin violating spreading widths.

We start with the nuclear Hamiltonian $H$ written in the form
\begin{equation}
\label{d8}
H=H_{0} \quad + \quad V_{S} \quad + \quad U_C \quad ,
\end{equation}
where the first term $H_{0}={\bf p}^{2}/2m+U_{S}(r)$ is the single-particle
Hamiltonian of nucleons moving in a single-particle 
strong potential $U_{S}(r)$.
We will use the harmonic oscillator form $U_S = \frac{m \omega^2}{2}r^2$. 
The term $V_{S}$ is the residual
two-body strong interaction.
The one-body Coulomb potential is denoted by $U_C$ and we use the
uniform density form:
\begin{eqnarray}
\label{EQ2}
U_{C}=3Ze^{2}/(2R)(1-r^{2}/(3R^2)), r\leq R
\nonumber\\
U_{C}=Ze^{2}/r, r>R, 
\end{eqnarray}
where $R$, $a$, and $r$ are the nuclear radius, 
diffuseness, and the radial variable correspondingly.

We will be interested below in the RPA renormalization of the
{\it off-diagonal} part of the Coulomb potential, $\tilde U_C$, defined
as the rest of $U_C$ after subtraction 
of $\bar U_C$ which is diagonal in the representation 
in which the single-particle Hamiltonian $H_0$ is diagonal
\begin{equation}
\label{OFF-DIAGONAL}
U_C \quad = \quad \tilde U_C  \quad + \quad \bar U_C.
\end{equation}
We now proceed \cite{FVPRC,FVPRC94}
by defining an anti-Hermitean operator $\hat A$ 
and perform the following unitary transformation.
The transformed Hamiltonian 
is:
\begin{eqnarray}
\label{d12}
H^* \quad = \quad e^{\hat A}H e^{-{\hat A}} \quad =
\qquad \qquad \qquad \qquad \qquad
\nonumber\\
H_{0} \quad + \quad V_{S} \quad + \quad \tilde U_C \quad +
\quad \bar U_C  \quad
+\quad [{\hat A},H_{0}] \quad
\nonumber\\
+ \quad [{\hat A},V_{S}],
\end{eqnarray}
where we have used the decomposition (\ref{OFF-DIAGONAL}) 
and neglected all terms above
the first order in the Coulomb interaction.
To obtain the effective one-particle 
off-diagonal Coulomb field, $\tilde U^*_C$, that results from the 
initial one, $\tilde U_C$,
due to the resudual strong interaction effects,
we should find  the
operator ${\hat A}$ in such a way that the 
single-particle off-diagonal Coulomb contribution
in $e^{\hat A}He^{-{\hat A}}$ will be cancelled. The last term 
in Eq.(\ref{d12}) is
a two-body operator. We introduce the decomposition: 
$[{\hat A},V_{S}] \equiv \langle [{\hat A},V_{S}] \rangle +
:[{\hat A},V_{S}]:$, where the
first is a single-particle term obtained by averaging over the nucleon
occupied states,
and the second two-body term, $:[{\hat A},V_{S}]:$, is the rest 
which yields zero under such averaging.

We can compensate the one-body Coulomb type field in the 
Hamiltonian $H^*$, if we require the 
operator $\hat A$ to satisfy the equation
\begin{equation}
\label{COMPENSATION}
\tilde U_C  \quad + \quad [{\hat A},H_{0}] \quad +
\quad \langle [{\hat A},V_{S}] \rangle =0.
\end{equation}
Then, the transformed Hamiltonian takes the form
\begin{equation}
\label{d15}
\tilde H = H_{0}  +  V_{S}  +  \bar U_C  + :[{\hat A},V_{S}]:
\end{equation}
where no single-particle off-diagonal Coulomb potential is present.
In this case, the effective off-diagonal Coulomb potential 
field can be found from the equation 
\begin{equation}
\label{NEW-FIELD}
\tilde U^*_C = - [ \hat A , H^0 ], 
\end{equation}
where the effects of the renormalization of the $\tilde U_C$ 
are incorporated into the operator $A$ \cite{FVPRC}.
In the language of the present approach, the operator $\hat A$ 
in Eq.(\ref{d12})
creates a small distortion of the nuclear density matrix 
when acting on the density matrix which contains no effects 
due to the residual interaction $V_S$.
Equations (\ref{COMPENSATION}),(\ref{NEW-FIELD}) 
can be seen to be equivalent to the main equation
of the Theory of Finite Fermi Systems 
\cite{MIGDAL}
\begin{equation}
\label{E-TFFS}
\tilde {\cal F}_{ik}^*  =  \tilde {\cal F}_{ik} + 
\sum_{\alpha \beta} (V_{S})_{\beta \alpha, i k}
\frac{n_{\beta} - n_{\alpha}}
{ \varepsilon_{\beta} - \varepsilon_{\alpha} - \omega}
\tilde {\cal F}_{ik}^*
\end{equation}
which relates an initial off-diagonal field $\tilde {\cal F}$
to its effective field $\tilde {\cal F}^*$, obtained via 
summation of the series of RPA diagrams involving
the residual interaction and the particle-hole propagator. ${\cal A}$
\cite{MIGDAL}. 
In Eq.(\ref{E-TFFS}), indices refer to the single-particle 
nucleon states which label the matrix elements of the fields
$\tilde {\cal F}$, $\tilde {\cal F}^*$, the interaction $V_S$ 
and the single-particle occupation probabilities
$n_i$ of the states with the energies $\varepsilon_i$.
In our case, the external (Coulomb) field is static,
so $\omega =0$.
Indeed, substituting Eq.(\ref{NEW-FIELD}) into Eq.(\ref{COMPENSATION})
and collecting the terms one obtains an equation analogous 
to Eq.(\ref{E-TFFS}) with $\tilde {\cal F}^*$ and $\tilde {\cal F}$
replaced by $\tilde {U}^*$ and $\tilde U$.
Equations (\ref{COMPENSATION}),(\ref{NEW-FIELD}) 
provide a convenient way to obtain  
analytical results for the renormalization of the potential
in the approximation which we develop below.

Up to now, we did not specify the explicit form of the 
residual strong interaction. 
To obtain results for the renormalization according to
Eqs.(\ref{COMPENSATION},\ref{NEW-FIELD}), 
one needs to specify $V_S$.
We use below the Landau-Migdal parametrization \cite{MIGDAL},\cite{BG} 
of the strong interaction
$V_S$,
\begin{equation}
\label{m10}
V_S({\bf r},
{\bf r}')=C\delta({\bf r}-{\bf r}')[f+f'{\bf \tau}{\bf \tau'}+
g{\bf \sigma}{\bf \sigma'}+g'{\bf \tau}{\bf \tau'}
{\bf \sigma}{\bf \sigma'}],
\end{equation}
which has been widely used for heavy nuclei  
(see \cite{MIGDAL},\cite{BG}).
One therefore considers the form (\ref{m10}) as a reliable
parametrization of $V_S$ that reproduces correctly the 
main properties of actual residual interaction.
The strengh constants $f$,$f'$,$g$, and $g'$ depend on the nuclear density. 
The relevant part of the interaction Eq.(\ref{m10}), $V_r$,
that is responsible for the renormalization
of the one-body off-diagonal Coulomb field is given by the first two terms
of Eq.(\ref{m10}) whose explicit form reads as
\begin{eqnarray}
\label{STRONG}
V_r({\bf r},{\bf r}')=C\delta({\bf r}-{\bf r}')[
f_{in}-(f_{ex}-f_{in})(\rho(r) - \rho_0)\rho_0^{-1} 
+
\nonumber\\
+(f'_{in}-(f'_{ex}-f'_{in})(\rho(r) - \rho_0)\rho_0^{-1}) 
({\bf \tau}{\bf \tau'})].
\end{eqnarray}
Here, $\rho(r)$ is the nuclear density, and 
$\rho_0$ denotes its value in the center
of nucleus. 
Here $C=300$ $MeV$ $fm^{3}$ is the universal
Migdal constant. The values of the strength parameters are \cite{MIGDAL}:
$f_{ex}=-1.95$, $f_{in}=-0.075$ $f'_{ex}=0.05$ $f'_{in}=0.675$.

From symmetry considerations, 
the operator $\hat A$ can be taken in the form:
\begin{equation}
\label{operatorA}
\hat A = x_p ((\vec{\nabla}_p \vec{r}_p) + (\vec{r}_p \vec{\nabla}_p))+
x_n ((\vec{\nabla}_n \vec{r}_n) + (\vec{r}_n \vec{\nabla}_n)),
\end{equation}
to match equation (\ref{COMPENSATION})
with some constants $x_p,x_n$ to be determined.

In the expression for $\langle [A , V_S] \rangle $ we have,
after some calculations,
\begin{equation}
\label{DOUBLE-COM}
\langle [A , V_S] \rangle 
=
2 \frac{C \rho_0}{|U_0|}
\left( S_p - R_p r^2_p + S_n - R_n r^2_n \right).
\end{equation}
The off-diagonal part of Eq.(\ref{DOUBLE-COM})
which is to be substituted into Eq.(\ref{COMPENSATION})
is given by the off-diagonal parts of the 
second and the fourth terms in (\ref{DOUBLE-COM}) 
involving the off-diagonal terms of $r^2_p$ and $r^2_n$
(denoted by  $\tilde r^2_p$ and $\tilde r^2_n$).
The following combinations are introduced:
\begin{eqnarray}
\label{CONSTANTS-S-R}
S_p=3U_0 \left(
x_p F_{pp} \frac{Z}{A} + x_n F_{pn} \frac{N}{A} -
x_p F'_{pp} \frac{Z}{A} - x_n F'_{pn} \frac{N}{A} \right)
\nonumber\\
R_p=m\omega^2 \left[ \frac{5}{2} \left(
x_p F_{pp} \frac{Z}{A} + x_n F_{pn} \frac{N}{A}\right) -
\gamma \left(
x_p F'_{pp} \frac{Z}{A} + x_n F'_{pn} \frac{N}{A} \right) \right]
\nonumber\\
S_n=3U_0 \left(
x_p F_{np} \frac{Z}{A} + x_n F_{nn} \frac{N}{A} -
x_p F'_{np} \frac{Z}{A} - x_n F'_{nn} \frac{N}{A} \right)
\nonumber\\
R_n=m\omega^2 \left[ \frac{5}{2} \left(
x_p F_{np} \frac{Z}{A} + x_n F_{nn} \frac{N}{A}\right) -
\gamma \left(
x_p F'_{np} \frac{Z}{A} + x_n F'_{nn} \frac{N}{A} \right) \right]
\end{eqnarray}
The new constants are defined as follows
\begin{eqnarray}
\label{NEW-CONSTANTS}
F_{pp}=F_{nn}=f_{ex}+f'_{in},  \quad 
F'_{pp}=F'_{nn}=-(f_{ex}-f_{in}+f'_{ex}-f'_{in}),
\nonumber\\
F_{pn}=F_{np}=f_{ex}-f'_{in},  \quad 
F'_{pn}=F'_{np}=-(f_{ex}-f_{in}-f'_{ex}+f'_{in}).
\end{eqnarray}
Now, substituting Eq.(\ref{DOUBLE-COM}) into Eq.(\ref{COMPENSATION})
and accouning for 
(\ref{CONSTANTS-S-R}),(\ref{NEW-CONSTANTS})
we obtain, after separating the similar operator structures,
the following system of equations:

\begin{eqnarray}
\label{COLL_SYSTEM}
\left( \begin{array}{cc}
1- \frac{Z}{2A} \frac{C \rho_0}{|U_0|}\left(\frac{5}{2}F_{pp}-
\gamma F'_{pp} \right) & -\frac{N}{2A}\frac{C \rho_0}{|U_0|}\left(\frac{5}{2}F_{pn}-\gamma F'_{pn} \right) 
\\ 
-\frac{Z}{2A}\frac{C \rho_0}{|U_0|}
\left(\frac{5}{2}F_{np}-\gamma F'_{np} \right) & 
1- \frac{N}{2A} \frac{C \rho_0}{|U_0|}\left(\frac{5}{2}F_{nn}-
\gamma F'_{nn} \right)  
\end{array} \right)
\left( \begin{array}{c}
x_p  \\ 
x_n  
\end{array} \right)
\quad = 
\left( \begin{array}{c}
\frac{q_p}{4 m \omega^2}  \\ 
0
\end{array} \right)
,
\end{eqnarray}
The solution for this simple system of linear equations
is found inverting the matrix in Eq.(\ref{COLL_SYSTEM});  
the value of the determinant is given by the formula
\begin{eqnarray}
\label{determinant}
D=
\left[1-\frac{Z}{2A}\frac{C\rho_0}{|U_0|}\left(\frac{5}{2}F_{pp}-\gamma F'_{pp}
\right)\right]
\left[
1-\frac{N}{2A}\frac{C\rho_0}{|U_0|}\left(\frac{5}{2}F_{nn}-\gamma F'_{nn}
\right)\right]
- \nonumber\\
-\frac{ZN}{4A^2} \left(\frac{C\rho_0}{|U_0|}\right)^2
\left(\frac{5}{2} F_{pn} - \gamma F'_{pn} \right)^2
\end{eqnarray}
Instead of the initial off-diagonal Coulomb field
\begin{equation}
\label{INITIAL}
\tilde U_{C}^{(p)} = -q \tilde r^2_p, \quad \tilde U_{C}^{(n)} = 0,
\end{equation}
using Eqs.(\ref{NEW-FIELD}),(\ref{operatorA})
and finding $x_p,x_n$ from the system of equation (\ref{COLL_SYSTEM}), 
we obtain now 
proton and neutron effective (renormalized) fields
which are given by the expressions:
\begin{eqnarray}
\label{REN-POT}
\tilde{U}^{*(p)}_{C} = 
-\frac{1 - \frac{N}{2A}\frac{C\rho_0}{|U_0|}
\left(\frac{5}{2} F_{nn} - \gamma F'_{nn} \right)}{D} q \tilde r^2_p ,
\nonumber\\
\tilde U^{*(n)}_{C} = 
-\frac{\frac{Z}{2A}\frac{C\rho_0}{|U_0|}
\left(\frac{5}{2} F_{np} - \gamma F'_{np} \right)}{D} q \tilde r^2_n ,
\end{eqnarray}
(the asterisk marks the renormalized quantities).
The effective field that acts on neutrons is non zero, 
it contains now the contributions of
the proton-neutron component of the residual strong interaction.
The sign of this contribution is opposite to that of the
proton field.

For the renormalization of the isovector part of the off-diagonal
Coulomb field, 
$\tilde U^{*(i)}_c \equiv \tilde{U}^{*(p)}_{C}-\tilde{U}^{*(n)}_{C}$,
that can be written as 
\begin{displaymath}
\tilde U^{*(i)}_c = (q^*_p-q^*_n) \tilde r^2 
\equiv \frac{1}{S} (q \tilde r^2) 
\end{displaymath}
we obtain from Eqs.(\ref{INITIAL},\ref{REN-POT}) the following expression:
\begin{eqnarray}
\label{SUPPRESSION}
\frac{1}{S}=
\frac{q^*_p - q^*_n}{q}=
\qquad \qquad \qquad \qquad \qquad \qquad
\nonumber\\
=\frac{
1-\frac{N}{2A}\frac{C\rho_0}{|U_0|}\left(\frac{5}{2}F_{nn}-\gamma F'_{nn}
\right)
-\frac{Z}{2A}\frac{C\rho_0}{|U_0|}\left(\frac{5}{2}F_{np}-\gamma F'_{np}
\right)
}{
\left[1-\frac{Z}{2A}\frac{C\rho_0}{|U_0|}\left(\frac{5}{2}F_{pp}-\gamma F'_{pp}
\right)\right]
\left[
1-\frac{N}{2A}\frac{C\rho_0}{|U_0|}\left(\frac{5}{2}F_{nn}-\gamma F'_{nn}
\right)\right]
-\frac{ZN}{4A^2} \left(\frac{C\rho_0}{|U_0|}\right)^2
\left(\frac{5}{2} F_{pn} - \gamma F'_{pn} \right)^2
}
\end{eqnarray}
This analytical result gives the value of the suppression
factor $S$ of the isospin violating part of the Coulomb field
as a function of the nuclear charge and mass, and the 
interaction constants. 
A simple numerical evaluation of Eq.(\ref{SUPPRESSION})
for the case $N \simeq Z$
and the values of the interaction strengths presented above 
(see Eq.(\ref{STRONG}))
gives the following
value of the suppression factor $S$
\begin{equation}
\label{S-THEOR}
S = 1.8 \pm 0.4
\end{equation}
where the uncertainty is mainly due to that
in the quantity $\gamma$.
The obtained value of the reduction factor 
squared 
$S^2 \simeq 3.4 \pm 1.4$
can be compared to the suppression 
factor that results from comparison of the experimental spreading
widths of the IAS resonances
with the values calculated using the unrenormalized
Coulomb potential:
\begin{equation}
S^2_{ exp} \sim 3-4
\end{equation}
Therefore, the
present theoretical result for the renormalization 
of the off-diagonal isovector Coulomb field can be considered 
as a microscopic theoretical explanation 
of reduced isospin symmetry breaking in nuclei.

The utilization of the above effective off-diagonal Coulomb potential 
is limited to the inside of the nucleus and to the states localized inside.
We should keep in mind that the reduction
derived above is applicable only to the $r^2$ part of Eq.(\ref{EQ2})
and is not applicable for the outside ($r>R$) part 
that goes as $\frac{1}{r}$.

The way in which our present result could be applied is illustrated
in the following two examples:

A. {\it Isospin mixing in ground states}. In order to calculate
isospin mixing we can use the potential $\tilde U^{*}_c$ 
in the expression \cite{AK}:
\begin{displaymath}
(\varepsilon^*)^2 =
\sum_{nlj} \frac{|\langle nlj |\tilde U^{*(i)}_c | n+1 l j \rangle |^2}
{(E_n - E_{n+1})^2}
\end{displaymath}
where $|nlj \rangle$ are single-paticle states and $|n+1 lj \rangle$
the corresponding radially excited s.p. states.
The denominator is the energy difference between 
the unperturbed single-particle states. 
The collective effects due to RPA correlations 
are taken into account in $\tilde U^{*(i)}_c$. We find that
\begin{displaymath}
(\varepsilon^*)^2 = \frac{\varepsilon^2}{S^2}
\end{displaymath}
where $\varepsilon$ is the amount of isospin mixing
in the single-particle model without the suppression effect calculated
in the present work.
We see that our model predicts a reduction in isospin mixing of about 
a factor $S^2 \sim 4$ compared to the pure single-particle model 
in which the renormalization of the off-diagonal Coulomb 
field due to the strong interaction is not taken into acount.

B.{\it Spreading width of isobaric analog resonances}.
In the single-particle approximation
the spreading width is given by \cite{AK,A,A-PLB}
\begin{displaymath}
\Gamma^{\downarrow}_{A} = 
- Im \langle A | \tilde U_c G^{+}_P \tilde U_c | A \rangle
\end{displaymath}
where $|A \rangle$ is the analog state, and $G^{+}_P$ is the optical potential model Green's function.
As already discussed this kind of calculation overstimates
the spreading widths of isobaric analog resonances by factors close to 5 
for standard optical potentials \cite{AK}. 
If we replace $\tilde U_c$ by $\tilde U^{*(i)}_c$
we find that the spreading widths are much closer to the experimental ones
and close to the results obtained in Ref.\cite{A-PLB} where
the collectivity of the IVM is taken into account.

We should stress that this reduction 
applies only to the spreading widths and not to the decay (escape) widths
$\Gamma^{\uparrow}_{A}$. The latter involves continuum wave-function which 
are not localized inside the nucleus and therefore $\Gamma^{\uparrow}_{A}$
depends on the $1/r$ part of $U_c$.
As mentioned above the reduction we find does not apply to $1/r$ 
and therefore it affects the escape width much less,
in agreement with the calculations in Ref.\cite{AB}.

Some of the reduction of isospin mixing occurs already in the 
Hartee-Fock (HF) approximation\cite{A}. The attractive proton-neutron
force couples the proton and neutron distributions reducing the difference 
in the two caused by the Coulomb force. In an unrestricted HF calculation
that allows for deformations and for charge-exchange modes one obtaines
in the wave fuctions components that can be expressed as a particles 
coupled to the various RPA states (including the IVM) of the core.
Thus in such extended HF calculation effects of the RPA will be included.
However the use of such unrestricted RPA is not practical and the use of
such scheme for the quantitites calculated in this work is limited for the 
following reasons. (a) In order to incorporate correctly isovector effects
in the HF one must include charge-exchange excitations, which is a 
difficult task. (b) The HF calculation in an odd A system involves 
additional approximations. (c) The usual HF approach in nuclei with a 
neutron excess
introduces spurious isospin mixing and the symmetry potential acts as an 
isospin breaking term\cite{A}. (d) The HF approximation cannot be used for 
unbound states.
  
To summarize, 
we have developed here an approximate analytical approach to
study the role played by the residual strong interaction 
in reducing the Coulomb matrix elements
violating isospin symmetry in nuclear states.
We calculated for the first time analytically
the RPA renormalization of the off-diagonal Coulomb potential 
starting from the nuclear Hamiltonian with the residual two-body 
strong nuclear forces.
When realistic nuclear forces (Landau-Migdal 
parametrization) is used our approach produces 
quenched Coulomb matrix elements that are capable of 
explaining the experimental isospin violating spreading widths.

This work is supported by 
the Grant for Basic Research of the Israeli Academy of Science.


\begin{thebibliography}{200}
\bibitem{AK}
N.Auerbach, J. H\"ufner, A.K. Kerman and C.M. Shakin,
Rev. Mod. Phys., {\bf 44}, 48 (1972).

\bibitem{MEKJAN}
A.Z.Mekjan, Phys. Rev. Lett., {\bf 25}, 888 (1970).

\bibitem{AB}
N.Auerbach and G.Bertsch, Phys. Lett. {\bf 43B}, 175 (1973).

\bibitem{A}
N.Auerbach,  Phys. Rep. {\bf C98}, 273 (1983).

\bibitem{AG}
N.Auerbach and N. van Giai, Phys. Lett. {\bf B72}, 289 (1978).

\bibitem{A-PLB}
N.Auerbach, Phys. Lett. {\bf 44B}, 241 (1973).

\bibitem{BM} A.Bohr and B.Mottelson, {\it Nuclear Structure} (Benjamin,
New York, 1969), Vol. 1.

\bibitem{HRW}
H.L.Harney, A.Richter and H.A.Weidenm\"uller,
Rev. Mod. Phys., {\bf 58}, 607 (1986).

\bibitem{MBES}
G.E.Mitchell, E.G.Bilpuch, P.M.Endt, and J.F. Shriner,
Phys.Rev.Lett., {\bf 61}, 1473 (1988).

\bibitem{GW}
T.Guhr and H.A. Weidenm\"uller, Ann.Phys. (N.Y.), {\bf 199}, 412 (1990).

\bibitem{HAW}
H.A. Weidenm\"uller, Nucl.Phys. {\bf A552}, 293c (1991).

\bibitem{BRODY}
T.A.Brody, J.Flores, J.B.french, P.A.Mello, A.Pandey, and
S.S.M.Wong, Rev. Mod.Phys. {\bf 53}, 385 (1981).

\bibitem{FVPRC}
V.V.Flambaum and O.K.Vorov,
Phys. Rev. {\bf C51}, 1521 (1995).

\bibitem{FVPRC94} 
V.V.Flambaum and O.K.Vorov, Phys.Rev. {\bf C49}, 1827 (1994).

\bibitem{MIGDAL}
A.B.Migdal, {\it Theory of Finite Fermi Systems and 
Applications to atomic Nuclei} (John Wiley \& Sons, New York, 1967). 

\bibitem{BG} G.E.Brown, Rev.Mod.Phys., {\bf 43},1 1971);
V.Klemt, S.A.Moszkowski, and J.Speth, Phys.Rev. {\bf C14}, 302 (1976);
J.Speth, E.Werner, and W.Wild, Phys.Rep. {\bf 33}, No.3, 127(1977);
G.Bertsch, D.Cha, and H.Toki, Phys.Rev. {\bf C24}, 533 (1981);
V.A.Khodel and E.E.Sapershtein, Phys.Rep. {\bf 92}, 183 (1982), 
K. Goeke, and J.Speth, Annu. Rev. Nucl. Part. Sci. {\bf 32}, 65
(1982); F.Osterfeld, Rev.Mod.Phys., {\bf 64}, 491 (1992), and references
therein. 

\end{thebibliography}
\end{document}